\documentclass[aps,prl,
  twocolumn,
  longbibliography
]{revtex4-2}
\usepackage{amsmath,amssymb,amsfonts,amsthm}
\usepackage{bm,mathrsfs,mathtools}
\usepackage[normalem]{ulem}
\usepackage{dsfont}
\usepackage{textcomp}
\usepackage{upgreek,textgreek}
\usepackage[
]{hyperref}
\usepackage{physics}
\usepackage{orcidlink}
\usepackage{graphicx}
\usepackage{float}
\usepackage{siunitx}
\usepackage{booktabs}
\usepackage{color}
\usepackage[dvipsnames]{xcolor}

\begin{document}
\title{Robustness of Off-Axis Electron Vortices in Nonuniform Magnetic Fields}

\author{Hui-Dong Huang\,\orcidlink{0009-0005-8959-8751}}
\thanks{These authors contributed equally to this work.}
\affiliation{\href{https://ror.org/00n9dn158}{Sino-French Institute of Nuclear Engineering and Technology}, \href{https://ror.org/0064kty71}{Sun Yat-Sen University}, Zhuhai 519082, China}

\author{Qi Meng\,\orcidlink{0009-0000-8564-7601}}%
\thanks{These authors contributed equally to this work.}
\affiliation{\href{https://ror.org/00n9dn158}{Sino-French Institute of Nuclear Engineering and Technology}, \href{https://ror.org/0064kty71}{Sun Yat-Sen University}, Zhuhai 519082, China}

\author{Zhi-Bin Wang\,\orcidlink{0000-0002-6812-7855}}%
\affiliation{\href{https://ror.org/00n9dn158}{Sino-French Institute of Nuclear Engineering and Technology}, \href{https://ror.org/0064kty71}{Sun Yat-Sen University}, Zhuhai 519082, China}

\author{Liang Lu\,\orcidlink{0000-0002-8497-0738}}%
\affiliation{\href{https://ror.org/00n9dn158}{Sino-French Institute of Nuclear Engineering and Technology}, \href{https://ror.org/0064kty71}{Sun Yat-Sen University}, Zhuhai 519082, China}

\author{Jian Chen\,\orcidlink{0000-0001-9807-489X}}%
\email[Contact author:~]{chenjian5@mail.sysu.edu.cn}
\affiliation{\href{https://ror.org/00n9dn158}{Sino-French Institute of Nuclear Engineering and Technology}, \href{https://ror.org/0064kty71}{Sun Yat-Sen University}, Zhuhai 519082, China}

\author{Li-Ping Zou\,\orcidlink{0000-0001-8976-9171}}%
\email[Contact author:~]{zoulp5@mail.sysu.edu.cn}
\affiliation{\href{https://ror.org/00n9dn158}{Sino-French Institute of Nuclear Engineering and Technology}, \href{https://ror.org/0064kty71}{Sun Yat-Sen University}, Zhuhai 519082, China}

\begin{abstract}
Rotational symmetry protects the topological charge of on-axis electron vortices but not of off-axis vortices. We identify an additional SU(1,1) dynamical invariant that guarantees conservation of their intrinsic orbital angular momentum within the near-axis approximation. First-principles simulations of an off-axis electron vortex traversing a Glaser lens confirm this prediction, establishing a robust transport mechanism in axisymmetric nonuniform magnetic fields.
\end{abstract}
\maketitle

\emph{Introduction.---}
Electron vortices carrying quantized orbital angular momentum (OAM) provide a controllable internal degree of freedom for free electrons~\cite{bliokh_2007_semiclassical,verbeeck_2010_production,bliokh_2017_theory,lloyd_2017_electron}, with versatile applications ranging from electron microscopy~\cite{juchtmans_2015_using}, quantum information~\cite{loffler_2023_quantum} and magnetic diagnostics~\cite{barrows_2022_3d} to high-energy physics~\cite{ivanov_2022_promises}. To date, their dynamics have been explored predominantly under strictly on-axis propagation
in axisymmetric magnetic fields, where rotational symmetry guarantees conservation of the
canonical OAM~\cite{bliokh_2012_electron,greenshields_2014_angular,zou_2021_general,
	melkani_2021_electron,karlovets_2026_angular,filina_2026_universal,filina_2026_effective}.

\begin{figure}
\centering
\includegraphics[width=0.7\linewidth]{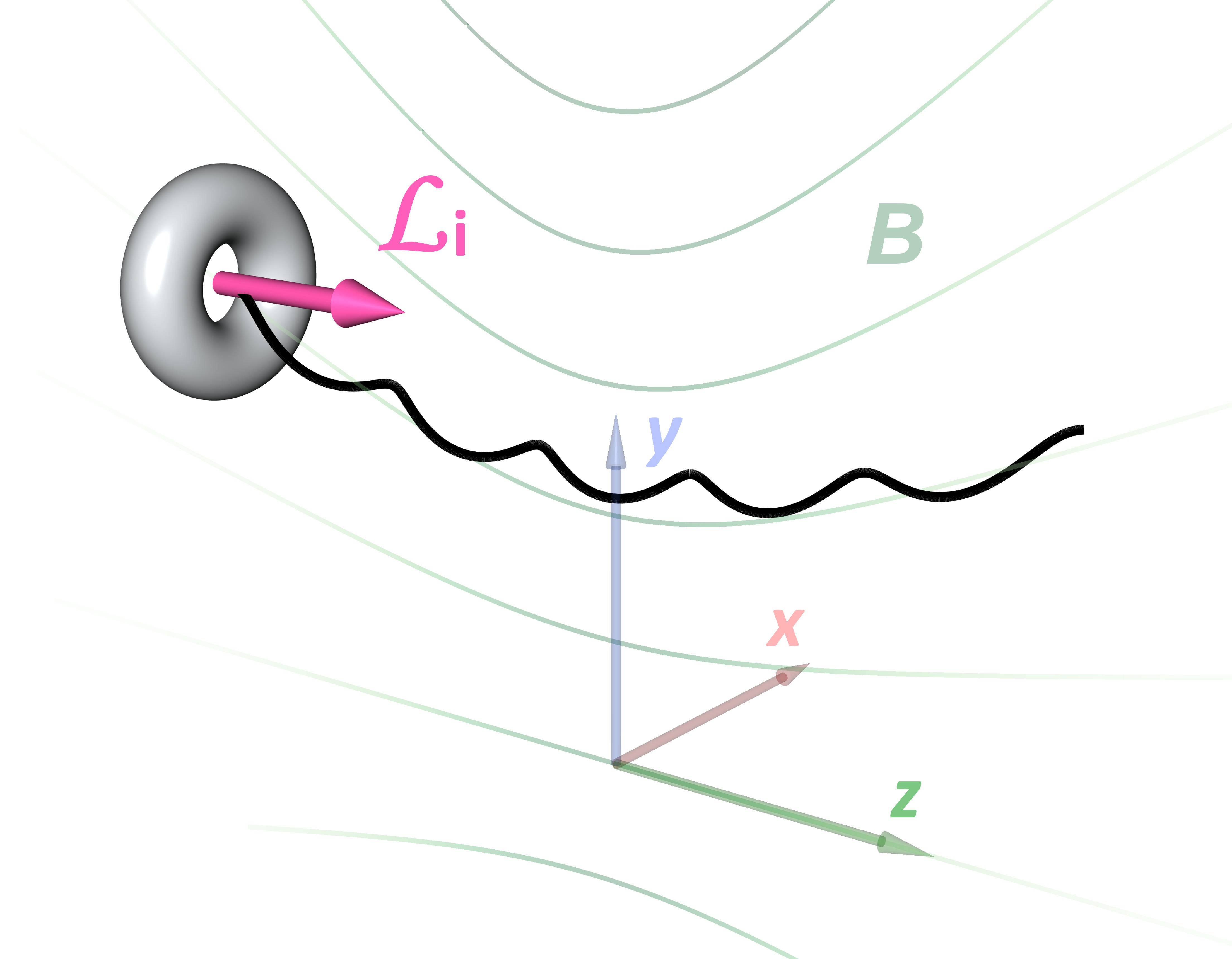}
\caption{An off-axis vortex wavepacket propagating through a magnetic lens. The centroid follows a helical trajectory that winds around the curved magnetic field lines, while the intrinsic OAM, defined relative to the centroid, is not protected by axial symmetry alone.
}
\label{fig:model}
\end{figure}

In practice, however, electron vortices inevitably propagate with some transverse
misalignment. In axisymmetric but nonuniform magnetic fields, such as those in
Glaser lenses~\cite{szilagyi_2012_electron}, the centroid then no longer coincides
with the symmetry axis (Fig.~\ref{fig:model}). Although axial symmetry still
guarantees conservation of the total canonical OAM, the intrinsic OAM, defined
relative to the instantaneous centroid~\cite{oneil_2002_intrinsic,greenshields_2015_parallel}, is not guaranteed by rotational symmetry
alone. Whether it survives off-axis propagation in such nonuniform fields remains
an open question. Dynamical invariants have previously been identified for on-axis vortex modes in nonuniform magnetic fields~\cite{melkani_2021_electron}, but those invariants constrain the transverse mode structure rather than the centroid motion; the intrinsic OAM of off-axis states was not addressed. More broadly, the conservation of extrinsic OAM lies outside the scope of Noether's theorem. It emerges from an algebraic structure of the near-axis
Hamiltonian rather than from a continuous symmetry, which may explain why it
has remained unnoticed.

In this Letter, we show that, within the paraxial approximation~\cite{bliokh_2017_theory},
the intrinsic OAM of an off-axis vortex is exactly conserved during propagation
through axisymmetric nonuniform magnetic fields, despite the absence of
centroid-frame symmetry protection. This conservation arises from two exact
constraints acting in concert: the axial symmetry of the external field fixes
the total canonical OAM, while a hidden SU(1,1) dynamical invariant, identified
with the extrinsic OAM, prevents any transfer of angular momentum into the
intrinsic part.

We confirm this prediction with first-principles numerical simulations of an
off-axis electron vortex traversing a Glaser lens. The simulations solve the
full three-dimensional Schr\"odinger equation without imposing the paraxial
approximation, while retaining only the near-axis vector potential, the latter
being far less restrictive than the paraxial condition. The results demonstrate that the wavepacket preserves both its vortex structure and intrinsic OAM over a wide range of propagation distances and transverse offsets, even under field gradients far steeper than typical experimental conditions. These findings establish a robust mechanism for vortex transport in realistic electron-optical systems,
where field inhomogeneity and beam misalignment are unavoidable.

\emph{Theory.---}
We consider an electron vortex in a magnetic field $\mathbf{B}$ rotationally symmetric
about the $z$-axis. Let $-e$ ($e>0$) be the electron charge. Near the axis, $\nabla\cdot\mathbf{B}=0$
implies an axially varying longitudinal field $B(z)$ accompanied by a weak radial
component at $\mathcal O(r_\perp)$, while off-axis corrections to $B_z$ enter at
$\mathcal O(r_\perp^2)$~\cite{zou_2024_production,melkani_2021_electron,filina_2026_universal}.
To leading order, the vector potential in the symmetric gauge is
\begin{equation}\label{eq:gauge}
	\mathbf A = \tfrac12 B(z)\,r_\perp \mathbf e_{\theta} .
\end{equation}

The Hamiltonian is $\hat{H} = \hat{\bm\pi}^2/2m$ with $\hat{\bm\pi}=\hat{\mathbf p}+e\mathbf A$.
The canonical OAM $\hat{\mathcal L}=(\hat{\mathbf r}\times\hat{\mathbf p})_z$ commutes with
$\hat{H}$, so $\langle\hat{\mathcal L}\rangle$ is exactly conserved by axial symmetry.
The intrinsic OAM, defined relative to the instantaneous centroid, decomposes as
\begin{equation}\label{eq:L_relation}
	\mathcal{L}_{\mathrm{i}} = \langle\hat{\mathcal L}\rangle - \mathcal{L}_{\mathrm{ext}},
	\quad
	\mathcal{L}_{\mathrm{ext}} = \langle\hat x\rangle\langle\hat p_y\rangle - \langle\hat y\rangle\langle\hat p_x\rangle.
\end{equation}
Hence $\mathcal{L}_{\mathrm{i}}$ evolves only through the extrinsic part $\mathcal{L}_{\mathrm{ext}}$, which is determined by the centroid's transverse motion. We note that $\mathcal L_{\mathrm{ext}}$
involves the canonical momentum, which is not gauge-invariant; as such it has no classical
counterpart and its conservation cannot be inferred from classical theorems.

Under the paraxial approximation $\psi=e^{ikz}\chi$ with $|\partial_z^2\chi|\ll k|\partial_z\chi|$
~\cite{bliokh_2017_theory}, the transverse dynamics obeys $i\hbar v\,\partial_z\chi=\hat H_{\mathrm{eff}}\chi$
($v=\hbar k/m$). With $\omega_L(z)=eB(z)/2m$, the effective Hamiltonian is
\begin{equation}\label{eq:Heff}
	\hat H_{\mathrm{eff}} = \frac{\hat{\mathbf p}_\perp^2}{2m} + \omega_L(z)\hat{\mathcal L} + \tfrac12 m\omega_L^2(z)\hat r_\perp^2 .
\end{equation}

We take the plane $z_0$ as the starting point of propagation, define
$\hat H_0\equiv\hat H_{\mathrm{eff}}(z_0)$, and work in the basis that diagonalizes
$\hat H_0$. At this plane, the vector potential \eqref{eq:gauge} yields a radial field
$B_r=-\tfrac12 r_\perp B'(z_0)$ while its linear $r_\perp$-dependence
ensures that $\hat H_0$ retains the algebraic form of a uniform-field
Landau Hamiltonian. It is diagonalized by ladder operators $\hat a\equiv\hat a(z_0)$, $\hat b\equiv\hat b(z_0)$, with $[\hat a,\hat a^\dagger]=[\hat b,\hat b^\dagger]=1$ and $[\hat a,\hat b]=[\hat a,\hat b^\dagger]=0$. These are Schr\"odinger-picture operators, with all $z$-dependence carried by the state.
Expectation values $\langle\cdot\rangle$ are understood to be taken with respect to
$|\chi(z)\rangle$.

In terms of $\hat a$ and $\hat b$,
\begin{equation}\label{eq:H0L}
	\hat H_0 = \hbar\omega_L(z_0)(2\hat a^\dagger\hat a+1),\qquad
	\hat{\mathcal L} = \hbar(\hat a^\dagger\hat a - \hat b^\dagger\hat b).
\end{equation}
The extrinsic OAM, defined in the Schr\"odinger picture, reads
\begin{equation}\label{eq:Lext_ab}
	\mathcal{L}_{\mathrm{ext}} = \hbar\left(|\langle\hat a\rangle|^2 - |\langle\hat b^\dagger\rangle|^2\right).
\end{equation}
The $z$-dependent part $\hat V(z)\equiv \hat H_{\mathrm{eff}}(z)-\hat H_0$ can be decomposed as
\begin{equation}\label{eq:Vdecomp}
	\hat V(z) = \Omega(z)\hat{\mathcal L} + \Lambda(z)\bigl(2\hat K_0 + \hat K_- + \hat K_+\bigr),
\end{equation}
with $\Omega=\omega_L(z)-\omega_L(z_0)$ and 
$\Lambda=\hbar[\omega_L^2(z)-\omega_L^2(z_0)]/2\omega_L(z_0)$.
Here
\begin{equation}\label{eq:SU11gen}
	\hat K_0=\tfrac12(\hat a^\dagger\hat a+\hat b^\dagger\hat b+1),\quad
	\hat K_+=\hat a^\dagger\hat b^\dagger,\quad
	\hat K_-=\hat a\hat b
\end{equation}
are the SU(1,1) generators, satisfying $[\hat K_0,\hat K_\pm]=\pm\hat K_\pm$ and $[\hat K_-,\hat K_+]=2\hat K_0$~\cite{gerry_1991_correlated}.

We now consider an off-axis vortex state at $z_0$ with 
$\mathcal{L}_{\mathrm{i}}(z_0)=\ell\hbar$, i.e. a state whose transverse centroid satisfies $\langle \mathbf{r}_\perp\rangle \neq 0$. 
According to the decomposition in Eq.~\eqref{eq:L_relation}, the extrinsic angular momentum $\mathcal{L}_{\mathrm{ext}}$ is determined by the centroid motion. 
For an on-axis state ($\langle \mathbf{r}_\perp\rangle = 0$), $\mathcal{L}_{\mathrm{ext}}$ vanishes identically; in uniform fields it may remain trivially constant~\cite{greenshields_2015_parallel}, and its dynamical role has often been overlooked. In the off-axis, nonuniform regime, however, the non-zero centroid ($\langle \mathbf{r}_\perp\rangle \neq 0$) makes $\mathcal{L}_{\mathrm{ext}}$ a genuine dynamical variable whose conservation must be examined directly.

The term $\hat H_0+\Omega(z)\hat{\mathcal L}$ can be removed by the unitary transformation
\begin{equation}
\hat U(z)=\exp\!\left[-\frac{i}{\hbar v}\int_{z_0}^z dz'\,
\bigl(\hat H_0+\Omega(z')\hat{\mathcal L}\bigr)\right].
\end{equation}
In the resulting interaction picture, $|\chi_I(z)\rangle=\hat U^\dagger(z)|\chi(z)\rangle$,
the evolution is governed by the pure SU(1,1) Hamiltonian
\begin{equation}
\hat H_I(z)=\Lambda(z)\bigl(2\hat K_0+e^{-i\Phi}\hat K_-+e^{i\Phi}\hat K_+\bigr),
\end{equation}
with
\begin{equation}
	\Phi(z)=\frac{2\omega_L(z_0)(z-z_0)}{v}.
\end{equation}

To find a conserved quantity, we switch to the Heisenberg picture within the
interaction frame. Defining $\hat a_I(z)=\hat U_I^\dagger(z)\hat a\,\hat U_I(z)$
and $\hat b_I^\dagger(z)=\hat U_I^\dagger(z)\hat b^\dagger\,\hat U_I(z)$, where
$\hat U_I(z)$ is the formal evolution operator generated by $\hat H_I(z)$, i.e.\,
$i\hbar v\,\partial_z\hat U_I=\hat H_I\hat U_I$ with $\hat U_I(z_0)=I$.
A symmetric rotation
\begin{equation}
\hat a_R=e^{-i\Phi/2}\hat a_I,\qquad \hat b_R^\dagger=e^{i\Phi/2}\hat b_I^\dagger
\end{equation}
then yields 
\begin{equation}\label{eq:a_eq}
i\hbar v\,\partial_z\begin{pmatrix}
\hat a_R\\
\hat b_R^\dagger
\end{pmatrix}
=\mathcal H(z)\begin{pmatrix}
	\hat a_R\\
	\hat b_R^\dagger
\end{pmatrix},	
\end{equation}
with
\begin{equation}
	\mathcal H(z)=
	\begin{pmatrix}
		\Delta(z) & \Lambda(z)\\
		-\Lambda(z) & -\Delta(z)
	\end{pmatrix},\quad
	\Delta(z)=\hbar\omega_L(z_0)+\Lambda(z).
\end{equation}
The expectation values $\langle\hat a_R\rangle_0$ and
$\langle\hat b_R^\dagger\rangle_0$, taken with respect to the initial
interaction-picture state $|\chi_I(z_0)\rangle$, obey the same equation (\ref{eq:a_eq}). Defining $\mathbf w=(\langle\hat a_R\rangle_0,\langle\hat b_R^\dagger\rangle_0)^T$,
one has $i\hbar v\,\partial_z\mathbf w=\mathcal H\mathbf w$. Taking the conjugate
transpose gives $-i\hbar v\,\partial_z\mathbf w^\dagger=\mathbf w^\dagger\mathcal H^\dagger$,
from which
\begin{equation}
i\hbar v\,\partial_z(\mathbf w^\dagger\sigma_z\mathbf w)
= \mathbf w^\dagger(\sigma_z\mathcal H-\mathcal H^\dagger\sigma_z)\mathbf w.
\end{equation}
Since $\mathcal H^\dagger\sigma_z=\sigma_z\mathcal H$, the right-hand side vanishes,
so the SU(1,1) norm $\mathbf w^\dagger\sigma_z\mathbf w=
|\langle\hat a_R\rangle_0|^2-|\langle\hat b_R^\dagger\rangle_0|^2$ is conserved.

To relate this conserved quantity to a physical observable, we return to the
Schr\"odinger picture, where $\mathcal{L}_{\mathrm{ext}}$ was expressed in terms
of $\hat a$ and $\hat b^\dagger$ in Eq.~\eqref{eq:Lext_ab}. Tracing the sequence
of transformations back to the original Schr\"odinger state, one finds that
$\langle\hat a_R\rangle_0$ and $\langle\hat b_R^\dagger\rangle_0$ differ from
$\langle\hat a\rangle$ and $\langle\hat b^\dagger\rangle$ by phases that drop out
in the norm. Hence
\begin{equation}
\mathbf w^\dagger\sigma_z\mathbf w
=|\langle\hat a\rangle|^2-|\langle\hat b^\dagger\rangle|^2
=\mathcal{L}_{\mathrm{ext}}/\hbar,
\end{equation}
which is therefore constant.

Like the classical adiabatic invariant $\mu=m v_\perp^2/(2B)$ (the magnetic moment associated with the particle's gyromotion, approximately conserved when the field varies slowly compared with the cyclotron period~\cite{northrop_1963_adiabatic}), $\mathcal{L}_{\mathrm{ext}}$ is exactly conserved within the paraxial approximation, beyond which the conservation is only approximate (as seen in the numerical simulations below), and likewise constrains the transverse motion while leaving the detailed dynamics undetermined. Unlike $\mu$, however, it is canonical and gauge-dependent, so it represents a purely quantum version of such an invariant. Although the near-axis Hamiltonian locally preserves the Landau eigenstructure at each $z$, an off-axis state feels transverse field components as it propagates, so its wavefunction can evolve in a complex way. The existence of this invariant protects the topological charge despite such complexity.

\begin{figure}
	\centering
	\includegraphics[width=0.94\linewidth]{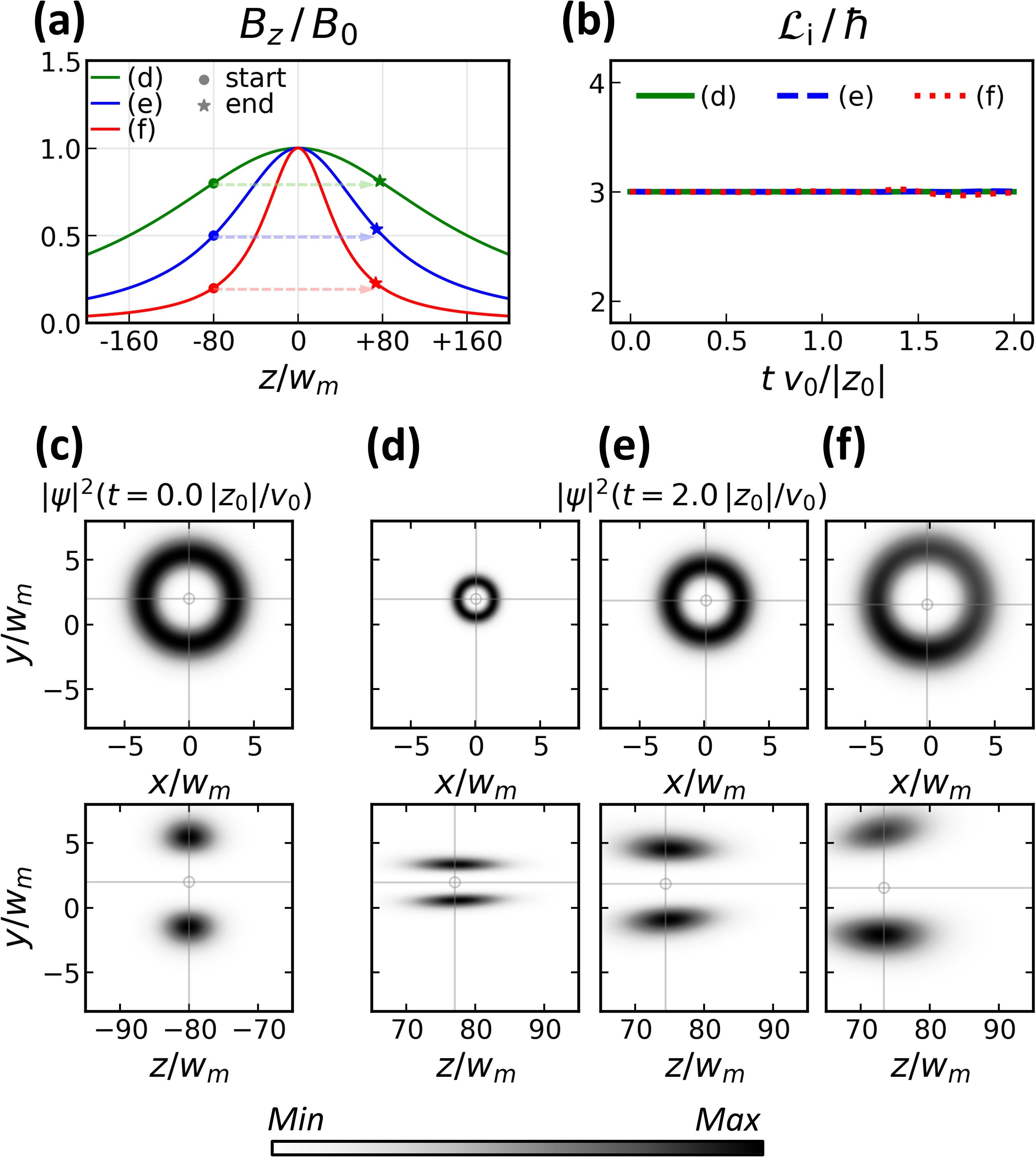}
	\caption{
		Dynamics of an off-axis vortex-electron wavepacket in Glaser lenses with different characteristic lengths $a$. Panel (a) shows the longitudinal magnetic-field profiles $B_z(z)$ for $a=160,80,40\,w_m$, and panel (b) the evolution of the intrinsic canonical OAM $\mathcal{L}_{\mathrm{i}}$. Panels (c--f) show the corresponding probability density $|\psi|^2$ at the initial time and at $t=2.0|z_0|/v_0$. The initial wavepacket has topological charge $\ell=+3$, characteristic scales $w_\perp=2\,w_m$ and $w_z=4\,w_m$, centroid position ${\bf r}_0=(0,+2,-80)\,w_m$, and axial speed $v_0=(160\,w_m)/(6T_c)$. Here $B_0=\SI{0.02}{T}$, corresponding to $w_m=\sqrt{2\hbar/|eB_0|}=\SI{0.256}{\micro\meter}$ and $T_c=2\pi m/|eB_0|=\SI{1.786}{\nano\second}$.
	}
	\label{fig:scan-a}
\end{figure}

\begin{figure}
	\centering
	\includegraphics[width=1\linewidth]{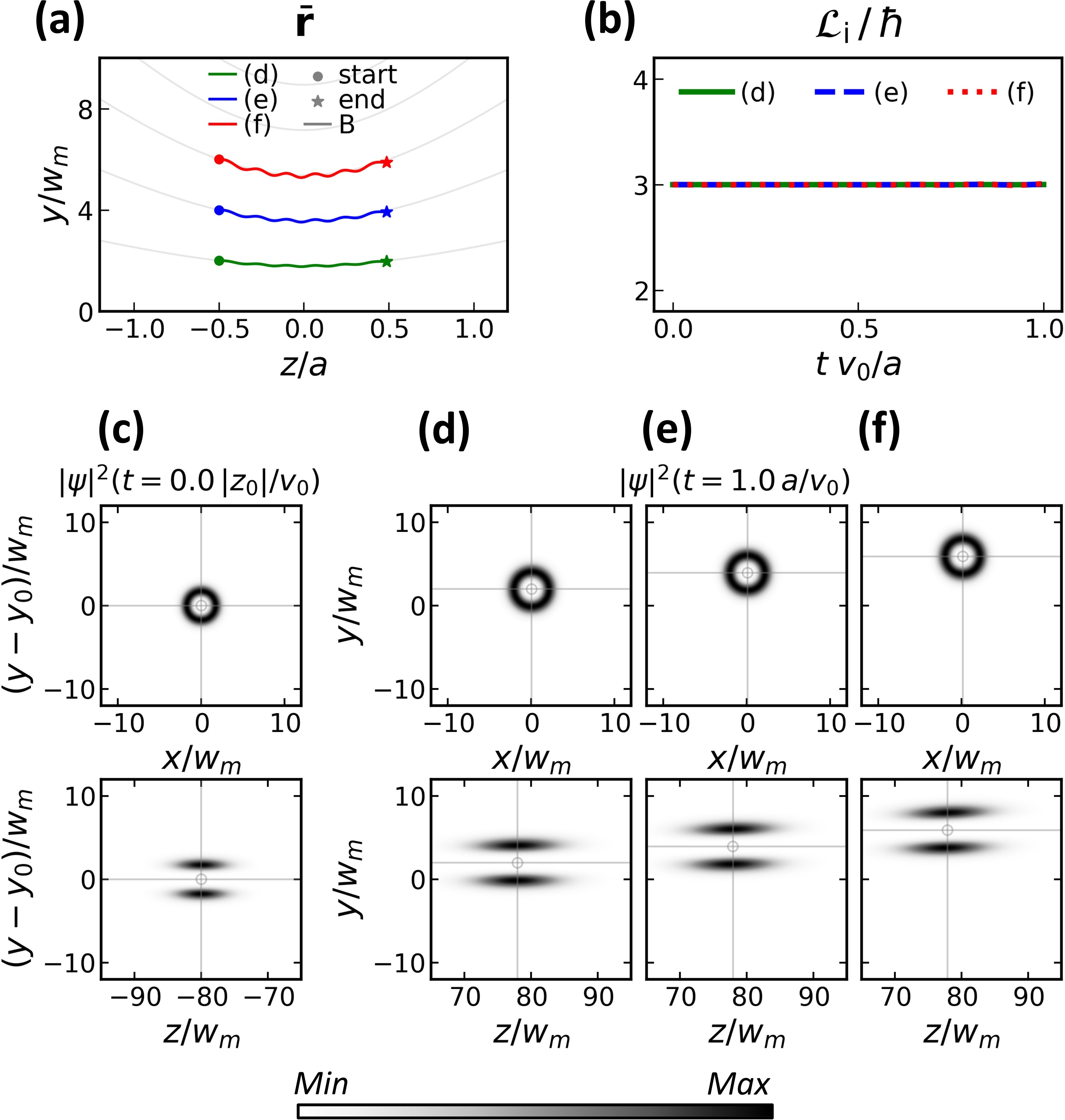}
	\caption{
		Dynamics of an off-axis vortex-electron wavepacket for different initial transverse offsets $y_0$. Panel (a) displays the centroid trajectories $\bar{\mathbf r}$ in the $z$--$y$ plane for $y_0=+2,+4,+6\,w_m$, while panel (b) shows the evolution of the intrinsic canonical OAM $\mathcal{L}_{\mathrm{i}}$.	Panels (c--f) show the corresponding probability density $|\psi|^2$ at the initial time and at $t=a/v_0$. The initial wavepacket has topological charge $\ell=+3$, characteristic scales $w_\perp=1\,w_m$ and $w_z=4\,w_m$, centroid position ${\bf r}_0=(0,y_0,-0.5\,a)$, and axial speed $v_0=a/(6T_c)$. The Glaser field is specified by $a=160\,w_m$ and $B_0=\SI{0.02}{T}$, yielding the characteristic scales $w_m=\SI{0.256}{\micro\meter}$ and $T_c=\SI{1.786}{\nano\second}$.
	}
	\label{fig:scan-d}
\end{figure}

\emph{Numerical simulation.---}
To verify the robustness of the vortex structure and intrinsic OAM during off-axis propagation through a Glaser magnetic lens with on-axis profile $B(z)=B_0(1+z^2/a^2)^{-1}$~\cite{szilagyi_2012_electron,khan_2021_quantum}, we solve the full Schr\"odinger equation. We employ an operator-splitting Fourier pseudo-spectral method combined with a semi-Lagrangian integration scheme~\cite{jin_2013_semilagrangian,caliari_2017_splitting,gutleb_2024_time}. In our simulations, a vortex-electron wavepacket traverses a Glaser lens, as sketched in Fig.~\ref{fig:model}. The initial state is a Laguerre-Gaussian vortex wavepacket with a Gaussian
longitudinal envelope~\cite{sizykh_2024_nonstationary}, whose toroidal
probability density $|\psi|^2$ is shown in Figs.~\ref{fig:scan-a}(c)
and \ref{fig:scan-d}(c).

We first examine the influence of magnetic non-uniformity, characterized by the length scale $a$, on the dynamics of an off-axis vortex electron, as shown in Fig.~\ref{fig:scan-a}. The parameter is scanned over $a\in[40,160]\,w_m$ while keeping the initial transverse offset fixed at $y_0=2\,w_m$. The resulting wavepacket snapshots at $t=2.0\,|z_0|/v_0$ reveal a clear dependence on the magnetic-field variation scale. For a sufficiently smooth field profile [large $a$, Fig.~\ref{fig:scan-a}(d)], the probability density exhibits transverse breathing while maintaining an almost ideal annular structure. As $a$ decreases, the sharper longitudinal field gradient gives rise to increasingly pronounced non-paraxial effects, manifested by a noticeable spatial tilt of the wavepacket in the $z$--$y$ plane [Fig.~\ref{fig:scan-a}(e--f)]. Throughout the parameter scan, however, the vortex structure remains intact (see Supplemental Material~\cite{sm} for the full time evolution). The evolution of the intrinsic canonical orbital angular momentum, $\mathcal{L}_{\mathrm{i}}$, is summarized in Fig.~\ref{fig:scan-a}(b). Although stronger field gradients lead to larger dynamical modulations during propagation, the deviation from its initial value, $\Delta\mathcal{L}_{\text{i}}=\mathcal{L}_{\text{i}}-\ell\hbar$ with $\ell=+3$, remains below $0.04\,\hbar$ over the full parameter range. Even in the most nonuniform case [(f): $a=40\,w_m$], the maximum relative deviation, $\varepsilon=\max|\Delta\mathcal{L}_{\text{i}}/\ell\hbar|$, is only $\sim1\%$.

We next investigate the influence of the initial transverse displacement $y_0$, as shown in Fig.~\ref{fig:scan-d}. Here the lens profile is fixed at $a=160\,w_m$, while the initial offset is varied over $y_0\in[+2,+6]\,w_m$. The centroid trajectories $\bar{\mathbf r}$ in the $z$--$y$ plane are shown in Fig.~\ref{fig:scan-d}(a), exhibiting the characteristic undulating cyclotron motion superimposed on the overall guiding motion along the curved magnetic field lines (gray curves). Although larger offsets expose the electron to stronger transverse field components, the transverse probability density consistently preserves its circular symmetry and annular vortex profile for all cases considered [Figs.~\ref{fig:scan-d}(d--f); see also Supplemental Material~\cite{sm}]. Likewise, the intrinsic OAM remains remarkably stable [Fig.~\ref{fig:scan-d}(b)], with the maximum relative deviation reaching only $\varepsilon\simeq0.3\%$ for the largest offset [(f): $y_0=+6\,w_m$], corresponding to $|\Delta\mathcal{L}_\text{i}|<0.01\,\hbar$ throughout the evolution.

To verify that these small residual variations are not numerical artifacts, we performed grid-refinement tests for the results shown in Figs.~\ref{fig:scan-a}(b) and \ref{fig:scan-d}(b). The percent-level residuals remain essentially unchanged under both coarser and finer spatial meshes (see Supplemental Material~\cite{sm}), indicating that the observed fluctuations in $\mathcal{L}_{\text{i}}$ originate from the underlying quantum dynamics rather than numerical discretization errors.

We further examine the dependence on the initial intrinsic OAM by fixing the lens profile at $a=160\,w_m$ and the initial offset at $y_0=+2\,w_m$, while changing the topological charge over $\ell\in[-5,+5]$. The maximum relative deviation remains below $\varepsilon\lesssim0.1\%$ for all values of $\ell$, with no clear scaling relationship with respect to $\ell$ (see Supplemental Material~\cite{sm}).

Taken together, these simulations confirm that neither magnetic-field non-uniformity nor off-axis beam injection significantly affects the intrinsic OAM. Both the vortex structure and the intrinsic canonical OAM remain remarkably robust during electron transport through the magnetic lens.

\emph{Discussion.---}
From a topological perspective, the vortex is characterized by the phase winding number around a closed contour $C$ enclosing the phase singularity at the beam centroid~\cite{lubk_2013_topoanaylsis},
$
\frac{1}{2\pi}\oint_C \boldsymbol{\nabla}_{\perp}\arg\phi\cdot d\boldsymbol{l}=\ell .
$
For the quadratic axisymmetric Hamiltonian \eqref{eq:Heff}, the wave function remains in the same azimuthal sector during propagation due to the hidden SU(1,1) dynamics, so the contour can be continuously transported without crossing a zero of the wave function. Consequently, the winding number cannot change, and the integer $\ell$ is a genuine conserved quantum number. In this model, the topological charge and the centroid-relative intrinsic OAM are represented by the same integer $\ell$, so intrinsic-OAM conservation is accompanied by exact preservation of the vortex topology.   

The small residual evolution of $\mathcal{L}_\mathrm{i}$ observed in our simulations
follows from the full three-dimensional quantum dynamics. From the Heisenberg
equations for $\hat{\mathbf r}$ and $\hat{\mathbf p}$, one obtains
\begin{equation}\label{eq:Lext}
	\begin{aligned}
		\dv{\mathcal{L}_{\mathrm{ext}}}{t}
		= &\;\; m\bigl[\langle\hat y\rangle\,\mathrm{Cov}(\hat\omega_L^2,\hat x)
		- \langle\hat x\rangle\,\mathrm{Cov}(\hat\omega_L^2,\hat y)\bigr] \\
		&+ \bigl[\langle\hat x\rangle\,\mathrm{Cov}(\hat\omega_L,\hat p_x)
		- \langle\hat p_x\rangle\,\mathrm{Cov}(\hat\omega_L,\hat x)\bigr] \\
		&+ \bigl[\langle\hat y\rangle\,\mathrm{Cov}(\hat\omega_L,\hat p_y)
		- \langle\hat p_y\rangle\,\mathrm{Cov}(\hat\omega_L,\hat y)\bigr],
	\end{aligned}
\end{equation}
where $\hat\omega_L = \omega_L(\hat z)$ and
$\mathrm{Cov}(\hat o_1,\hat o_2)\equiv \langle\hat o_1\hat o_2\rangle-\langle\hat o_1\rangle\langle\hat o_2\rangle$.
As $\langle\hat{\mathcal L}\rangle$ is conserved by axial symmetry, the residual
evolution of $\mathcal{L}_{\mathrm{i}}$ seen in our simulations must originate from
the right-hand side of Eq.~(\ref{eq:Lext}).

To see when these covariance terms become appreciable, expand
$\omega_L(\hat z) \simeq \omega_L(z_c) + \omega_L'(z_c)(\hat z - z_c)$
around the centroid $z_c \equiv \langle\hat z\rangle$. To leading order,
$\mathrm{Cov}(\hat\omega_L,\hat o) \simeq \omega_L'\,\mathrm{Cov}(\hat z,\hat o)$,
$\mathrm{Cov}(\hat\omega_L^2,\hat o) \simeq 2\omega_L\omega_L'\,\mathrm{Cov}(\hat z,\hat o)$
for $\hat o = \hat x, \hat y, \hat p_x, \hat p_y$.
For a wavepacket of longitudinal extent $w_z$ and transverse extent $w_\perp$,
$\mathrm{Cov}(\hat z,\hat r_{\perp}) \sim w_z w_\perp$ and
$\mathrm{Cov}(\hat z,\hat p_{\perp}) \sim w_z\,\hbar/w_\perp$, as these covariances
measure the characteristic longitudinal--transverse correlations within the
wavepacket.

Inserting these estimates into Eq.~(\ref{eq:Lext}), each of the six terms
scales as $|\omega_L'| w_z \hbar$ (the $\omega_L^2$ terms carry an extra factor
$m\omega_L w_\perp^2/\hbar \sim (w_\perp/w_m)^2$, which is $O(1)$ since
$w_\perp \sim w_m$ in our simulations).
With $\mathcal{L}_{\mathrm{ext}} \sim \hbar$ and dividing by the natural frequency scale $\omega_L$ yields the
dimensionless control parameter
\begin{equation}
\frac{|d\mathcal{L}_{\mathrm{ext}}/dt|/\mathcal{L}_{\mathrm{ext}}}{\omega_L}
\sim \frac{|\omega_L'| w_z}{\omega_L} \sim \frac{B' w_z}{B},
\end{equation}
which measures the fractional field variation across the wavepacket.
When $B' w_z/B \ll 1$, the field is nearly uniform and the covariance terms
are frozen out; when it approaches unity, the wavepacket resolves the
inhomogeneity and $\mathcal{L}_{\mathrm{ext}}$ can evolve appreciably.
For the Glaser field $B(z)=B_0(1+z^2/a^2)^{-1}$, $B'/B \sim 1/a$, so the
ratio reduces to $w_z/a$.

In our simulations this ratio reaches $w_z/a \lesssim 10^{-1}$
(the field changes by $\sim 10\%$ over the wavepacket), a regime far more extreme
than typical electron-optical conditions, chosen deliberately to expose any
beyond-paraxial corrections. Even under this strict test,
$\mathcal{L}_{\mathrm{i}}$ deviates by only a few percent. In practical settings,
where $w_z \ll B/B'$, the field is effectively uniform across the wavepacket,
the covariances are negligible, and $\mathcal{L}_{\mathrm{i}}$ is conserved.
Importantly, the theoretical results rely only on the near-axis vector potential
of Eq.~(\ref{eq:gauge}), not on the specific Glaser profile used in the numerical simulation. The mechanism thus applies generally to axisymmetric magnetic systems described by the same near-axis
model, including magnetic mirrors and solenoids~\cite{filina_2026_universal,
	zou_2024_production,sizykh_2024_nonstationary}.

Beyond electron optics, the demonstrated robustness also supports emerging proposals for off-axis vortex-particle scattering employing transverse beam-center offsets as tunable impact parameters~\cite{yang_2026_universal}, by preserving the intrinsic OAM during transport through axisymmetric nonuniform magnetic fields.

\emph{Summary.---}
We have shown that off-axis vortex electrons possess a previously unrecognized robustness in nonuniform axisymmetric magnetic fields. Although rotational symmetry alone does not protect the intrinsic OAM defined with respect to the instantaneous centroid, its conservation follows from the combined action of this geometric symmetry and a hidden dynamical invariant associated with the SU(1,1) algebra within the near-axis approximation. Numerical simulations of vortex-electron propagation through Glaser magnetic lenses verify this prediction and demonstrate stable transport of the vortex structure under magnetic-field inhomogeneity and beam misalignment. The mechanism is expected to extend to a broad class of axisymmetric electron-optical systems.

\emph{Acknowledgments.---}
We thank Pengming Zhang for valuable comments. The work was supported by the National Key R\&D Program of China (No. 2024YFE0109802), the National Natural Science Foundation of China (Grant No. 12175320), and the Guangdong Basic and Applied Basic Research Foundation (Grant No.2026A1515010999).


\end{document}


\title{Supplemental Material for\\“Robustness of Off-Axis Electron Vortices in Nonuniform Magnetic Fields”}

\author{Hui-Dong Huang\,\orcidlink{0009-0005-8959-8751}}
\thanks{These authors contributed equally to this work.}
\affiliation{\href{https://ror.org/00n9dn158}{Sino-French Institute of Nuclear Engineering and Technology}, \href{https://ror.org/0064kty71}{Sun Yat-Sen University}, Zhuhai 519082, China}

\author{Qi Meng\,\orcidlink{0009-0000-8564-7601}}%
\thanks{These authors contributed equally to this work.}
\affiliation{\href{https://ror.org/00n9dn158}{Sino-French Institute of Nuclear Engineering and Technology}, \href{https://ror.org/0064kty71}{Sun Yat-Sen University}, Zhuhai 519082, China}

\author{Zhi-Bin Wang\,\orcidlink{0000-0002-6812-7855}}%
\affiliation{\href{https://ror.org/00n9dn158}{Sino-French Institute of Nuclear Engineering and Technology}, \href{https://ror.org/0064kty71}{Sun Yat-Sen University}, Zhuhai 519082, China}

\author{Liang Lu\,\orcidlink{0000-0002-8497-0738}}%
\affiliation{\href{https://ror.org/00n9dn158}{Sino-French Institute of Nuclear Engineering and Technology}, \href{https://ror.org/0064kty71}{Sun Yat-Sen University}, Zhuhai 519082, China}

\author{Jian Chen\,\orcidlink{0000-0001-9807-489X}}%
\email[Contact author:~]{chenjian5@mail.sysu.edu.cn}
\affiliation{\href{https://ror.org/00n9dn158}{Sino-French Institute of Nuclear Engineering and Technology}, \href{https://ror.org/0064kty71}{Sun Yat-Sen University}, Zhuhai 519082, China}

\author{Li-Ping Zou\,\orcidlink{0000-0001-8976-9171}}%
\email[Contact author:~]{zoulp5@mail.sysu.edu.cn}
\affiliation{\href{https://ror.org/00n9dn158}{Sino-French Institute of Nuclear Engineering and Technology}, \href{https://ror.org/0064kty71}{Sun Yat-Sen University}, Zhuhai 519082, China}

\maketitle

\tableofcontents

\clearpage

\section{Landau-level algebra}
Starting from the canonical commutation relations
\begin{equation}
[\hat r_i,\hat p_j]=i\hbar\delta_{ij},\qquad
[\hat r_i,\hat r_j]=[\hat p_i,\hat p_j]=0,
\end{equation}
and kinetic momentum commutator
\begin{equation}
[\hat{\pi}_x,\hat{\pi}_y]=-{\rm i}\hbar eB_0,
\end{equation}
we can construct the cyclotron and guiding-center coordinates as
\begin{equation}
\hat{\boldsymbol\eta}=-\frac{1}{eB_0}\mathbf e_z\times\hat{\boldsymbol\pi},
\qquad
\hat{\mathbf R}=\hat{\mathbf r}_{\perp}-\hat{\boldsymbol\eta},
\end{equation}
which, in components,
\begin{equation}
\hat\eta_x=\frac{1}{eB_0}\hat\pi_y,\quad
\hat\eta_y=-\frac{1}{eB_0}\hat\pi_x,\quad
\hat X=\hat x-\hat\eta_x,\quad
\hat Y=\hat y-\hat\eta_y,
\end{equation}
satisfy
\begin{equation}
[\hat\eta_x,\hat\eta_y]= -\frac{i\hbar}{eB_0},\qquad
[\hat X,\hat Y]= \frac{i\hbar}{eB_0},\qquad
[\hat{\boldsymbol\eta},\hat{\mathbf R}]=0.
\end{equation}

We now introduce two pairs of ladder operator
\begin{equation}
\hat a=\sqrt{\frac{eB_0}{2\hbar}}(\hat\eta_x-i\hat\eta_y),\quad
\hat a^\dagger=\sqrt{\frac{eB_0}{2\hbar}}(\hat\eta_x+i\hat\eta_y),\quad
\hat b=\sqrt{\frac{eB_0}{2\hbar}}(\hat X+i\hat Y),\quad
\hat b^\dagger=\sqrt{\frac{eB_0}{2\hbar}}(\hat X-i\hat Y).
\end{equation}
They satisfy:
\begin{equation}
[\hat a,\hat a^\dagger]=1,\qquad
[\hat b,\hat b^\dagger]=1,\qquad
[\hat a,\hat b]=[\hat a,\hat b^\dagger]=0.
\end{equation}

Inverting these relations gives
\begin{equation}
\hat\eta_x=\sqrt{\frac{\hbar}{2eB_0}}(\hat a+\hat a^\dagger),\quad
\hat\eta_y=i\sqrt{\frac{\hbar}{2eB_0}}(\hat a-\hat a^\dagger),\quad
\hat X=\sqrt{\frac{\hbar}{2eB_0}}(\hat b+\hat b^\dagger),\quad
\hat Y=-i\sqrt{\frac{\hbar}{2eB_0}}(\hat b-\hat b^\dagger).
\end{equation}

Hence the position operator decomposes as
\begin{equation}
\hat{\mathbf r}_{\perp}=\hat{\boldsymbol\eta}+\hat{\mathbf R},
\end{equation}
i.e.
\begin{equation}
\hat x=\sqrt{\frac{\hbar}{2eB_0}}(\hat a+\hat a^\dagger+\hat b+\hat b^\dagger),\qquad
	\hat y=i\sqrt{\frac{\hbar}{2eB_0}}\left(\hat a-\hat a^\dagger-\hat b+\hat b^\dagger\right).
\end{equation}

The kinetic momentum depends only on the cyclotron sector,
\begin{equation}
\hat\pi_x=-eB_0\hat{\eta}_y=-i\sqrt{\frac{\hbar eB_0}{2}}(\hat a-\hat a^\dagger),\qquad
\hat\pi_y=eB_0\hat{\eta}_x=\sqrt{\frac{\hbar eB_0}{2}}(\hat a+\hat a^\dagger).
\end{equation}

The canonical momentum $\hat{\mathbf p}=\hat{\boldsymbol\pi}-e\mathbf A(\hat{\mathbf r})$ becomes
\begin{equation}
\hat p_x=\hat\pi_x+\frac{eB_0}{2}\hat{y}=
-\frac{i}{2}\sqrt{\frac{\hbar eB_0}{2}}
\left(\hat a-\hat a^\dagger+\hat b-\hat b^\dagger\right),\quad
\hat p_y=\hat\pi_y-\frac{eB_0}{2}\hat{x}=
\frac{1}{2}\sqrt{\frac{\hbar eB_0}{2}}
\left(\hat a+\hat a^\dagger-\hat b-\hat b^\dagger\right).
\end{equation}

The Landau Hamiltonian
\begin{equation}
\hat H_0=\frac{\hat{\boldsymbol\pi}^2}{2m}
\end{equation}
reduces to
\begin{equation}
\hat H_0=\hbar\omega_{L,0}\left(2\hat a^\dagger\hat a+1\right),
\end{equation}
showing that it depends only on the cyclotron sector, while $n_b$ labels the degeneracy.

The canonical orbital angular momentum
\begin{equation}
\hat{\mathcal L}=\hat x\hat p_y-\hat y\hat p_x
\end{equation}
takes the compact form
\begin{equation}
\hat{\mathcal L}=\hbar(\hat a^\dagger\hat a-\hat b^\dagger\hat b),
\end{equation}
so that eigenstates $|n_a,n_b\rangle$ carry quantum number $\ell=n_a-n_b$.

The extrinsic orbital angular momentum can be evaluated as follows:
\begin{equation}
\begin{aligned}
\mathcal{L}_{\text{ext}}&=\langle \hat{x}\rangle\langle \hat{p}_y\rangle-\langle \hat{y}\rangle \langle \hat{p}_x\rangle\\
&=\sqrt{\frac{\hbar}{2eB_0}}\left(\langle \hat{a}\rangle +\langle \hat{a}^{\dagger}\rangle+\langle \hat{b}\rangle+\langle \hat{b}^{\dagger}\rangle\right)\times \frac{1}{2}\sqrt{\frac{\hbar eB_0}{2}}\left(\langle \hat{a}\rangle +\langle \hat{a}^{\dagger}\rangle-\langle \hat{b}\rangle-\langle \hat{b}^{\dagger}\rangle\right)\\
&\qquad -i\sqrt{\frac{\hbar}{2eB_0}}\left(\langle \hat{a}\rangle -\langle \hat{a}^{\dagger}\rangle-\langle \hat{b}\rangle+\langle \hat{b}^{\dagger}\rangle\right)\times \left(-\frac{i}{2}\right)\sqrt{\frac{\hbar eB_0}{2}}\left(\langle \hat{a}\rangle -\langle \hat{a}^{\dagger}\rangle+\langle \hat{b}\rangle-\langle \hat{b}^{\dagger}\rangle\right)\\
&=\hbar \left(\langle \hat{a}\rangle\langle \hat{a}^{\dagger}\rangle-\langle \hat{b}\rangle\langle \hat{b}^{\dagger}\rangle\right)\\
&=\hbar\left(|\langle \hat{a}\rangle|^2-|\langle \hat{b}^{\dagger}\rangle|^2\right)
\end{aligned}
\end{equation}

Finally,
\begin{equation}
\hat{r}_{\perp}^2
=\hat{x}^2+\hat{y}^2
=\frac{2\hbar}{eB_0}\left(\hat{a}^{\dagger}\hat{a}+\hat{b}^{\dagger}\hat{b}+\hat{a}\hat{b}+\hat{a}^{\dagger}\hat{b}^{\dagger}+1\right).
\end{equation}

\clearpage

\section{Effective Hamiltonian and SU(1,1) structure}

\subsection{Decomposition of the effective Hamiltonian}

The paraxial effective Hamiltonian for an axisymmetric nonuniform magnetic field reads
\begin{equation}
	\hat H_{\mathrm{eff}}(z)=\frac{\hat{\mathbf p}_\perp^2}{2m}+\omega_L(z)\hat{\mathcal L}+\frac12 m\omega_L^2(z)\,\hat r_\perp^2,
	\label{eq:Heff}
\end{equation}
with $\omega_L(z)=eB(z)/2m$ and $\hat{\mathcal L}=\hat x\hat p_y-\hat y\hat p_x$.
We fix a reference plane $z_0$ and write $\omega_L(z)=\omega_L(z_0)+\Omega(z)$, with $\Omega(z)\equiv\omega_L(z)-\omega_L(z_0)$.
The $z_0$ terms are regrouped as
\begin{equation}
	\frac{\hat{\mathbf p}_\perp^2}{2m}+\omega_L(z_0)\hat{\mathcal L}+\frac12 m\omega_L^2(z_0)\,\hat r_\perp^2 = \hat H_0,
\end{equation}
where $\hat{H}_0$ is diagonal in the Fock basis $|n_a,n_b\rangle$.
Substituting this decomposition into \eqref{eq:Heff} yields
\begin{align}
	\hat H_{\mathrm{eff}}(z) &=
	\hat H_0+\Omega(z)\hat{\mathcal L}
	+\frac12 m\bigl[\omega_L^2(z)-\omega_L^2(z_0)\bigr]\hat r_\perp^2 .
	\label{eq:Heffsplit}
\end{align}

\subsection{SU(1,1) structure}

The squared transverse radius expressed in Landau ladder operators is
\begin{equation}
	\hat r_\perp^2 = \frac{2\hbar}{eB_0}\bigl(2\hat K_0+\hat K_-+\hat K_+\bigr),
\end{equation}
where $B_0=B(z_0)$ and the SU(1,1) generators are
\begin{equation}
	\hat K_0=\tfrac12(\hat a^\dagger\hat a+\hat b^\dagger\hat b+1),\qquad
	\hat K_+=\hat a^\dagger\hat b^\dagger,\qquad
	\hat K_-=\hat a\hat b,
\end{equation}
satisfying $[\hat K_0,\hat K_\pm]=\pm\hat K_\pm$ and $[\hat K_-,\hat K_+]=2\hat K_0$.
Using $eB_0 = 2m\omega_L(z_0)$, the perturbation becomes
\begin{align}
	\frac12 m\bigl[\omega_L^2(z)-\omega_L^2(z_0)\bigr]\hat r_\perp^2
	&= \frac12 m\bigl[\omega_L^2(z)-\omega_L^2(z_0)\bigr]
	\frac{2\hbar}{2m\omega_L(z_0)}\bigl(2\hat K_0+\hat K_-+\hat K_+\bigr) \nonumber\\
	&= \Lambda(z)\bigl(2\hat K_0+\hat K_-+\hat K_+\bigr),
	\label{eq:Vsu11}
\end{align}
where
\begin{equation}
	\Lambda(z)\equiv\frac{\hbar\bigl[\omega_L^2(z)-\omega_L^2(z_0)\bigr]}{2\omega_L(z_0)}.
\end{equation}

The canonical OAM commutes with all SU(1,1) generators:
\begin{equation}
	[\hat{\mathcal L},\hat K_0]=0,\qquad
	[\hat{\mathcal L},\hat K_\pm]=0.
\end{equation}
The first follows from $[\hat{\mathcal L},\hat a^\dagger\hat a]=[\hat{\mathcal L},\hat b^\dagger\hat b]=0$. For $\hat K_+=\hat a^\dagger\hat b^\dagger$,
$[\hat{\mathcal L},\hat K_+]=[\hat{\mathcal L},\hat a^\dagger]\hat b^\dagger+\hat a^\dagger[\hat{\mathcal L},\hat b^\dagger]=(\hbar\hat a^\dagger)\hat b^\dagger+\hat a^\dagger(-\hbar\hat b^\dagger)=0$, and similarly for $\hat K_-$.
Thus the complete $z$-dependent part takes the form
\begin{equation}
	\hat V(z)=\Omega(z)\hat{\mathcal L}+\Lambda(z)\bigl(2\hat K_0+\hat K_-+\hat K_+\bigr).
	\label{eq:Vfinal}
\end{equation}

\subsection{Factorization of the unitary transformation}

The term $\hat H_0+\Omega(z)\hat{\mathcal L}$ can be removed by
\begin{equation}
	\hat U(z)=\exp\!\Big[-\frac{i}{\hbar v}\int_{z_0}^z dz'\,\bigl(\hat H_0+\Omega(z')\hat{\mathcal L}\bigr)\Big].
\end{equation}
Since $[\hat H_0,\hat{\mathcal L}]=0$, the exponential factorizes:
\begin{equation}
	\hat U(z)=\exp\!\Big(-\frac{i}{\hbar v}\hat H_0(z-z_0)\Big)\;
	\exp\!\Big(-\frac{i}{\hbar v}\int_{z_0}^z dz'\,\Omega(z')\hat{\mathcal L}\Big)
	\equiv \hat U_0(z)\,\hat U_L(z).
	\label{eq:Ufactor}
\end{equation}

The required commutators are
\begin{align}
	[\hat H_0,\hat a] &= -2\hbar\omega_L(z_0)\hat a, \qquad
	[\hat H_0,\hat b^\dagger] = 0,\\
	[\hat{\mathcal L},\hat a] &= -\hbar\hat a, \qquad
	[\hat{\mathcal L},\hat b^\dagger] = -\hbar\hat b^\dagger.
\end{align}

A BCH calculation gives the combined transformation
\begin{align}
	\hat U^\dagger(z)\,\hat a\,\hat U(z) &= \hat a\,e^{-i[\Phi(z)+\Phi_1(z)]},\\
	\hat U^\dagger(z)\,\hat b^\dagger\,\hat U(z) &= \hat b^\dagger\,e^{-i\Phi_1(z)},
\end{align}
where
\begin{equation}
	\Phi(z)=\frac{2\omega_L(z_0)(z-z_0)}{v},\qquad
	\Phi_1(z)=\frac{1}{v}\int_{z_0}^z dz'\,\Omega(z').
\end{equation}

\subsection{Transformation of the SU(1,1) generators}

Since $[\hat{\mathcal L},\hat K_\pm]=0$, the $\Phi_1$ phases cancel in $\hat K_\pm$:
\begin{align}
	\hat U^\dagger \hat K_0 \hat U &= \hat K_0,\\
	\hat U^\dagger \hat K_\pm \hat U &= \hat K_\pm\,e^{\pm i\Phi(z)}.
	\label{eq:SUtrans}
\end{align}

\subsection{Interaction picture}

The Schr\"odinger-picture state satisfies $i\hbar v\,\partial_z|\chi(z)\rangle=\hat H_{\mathrm{eff}}(z)|\chi(z)\rangle$.
Defining $|\chi_I(z)\rangle=\hat U^\dagger(z)|\chi(z)\rangle$, a standard calculation gives
\begin{equation}
	i\hbar v\,\partial_z|\chi_I(z)\rangle=\hat H_I(z)|\chi_I(z)\rangle,
\end{equation}
with
\begin{equation}
	\hat H_I(z)=\Lambda(z)\bigl(2\hat K_0+e^{-i\Phi}\hat K_-+e^{i\Phi}\hat K_+\bigr).
	\label{eq:HI}
\end{equation}

\subsection{Heisenberg equations in the interaction frame}

Introducing $\hat U_I(z)$ via $i\hbar v\,\partial_z\hat U_I=\hat H_I\hat U_I$, $\hat U_I(z_0)=I$, the Heisenberg operators are
\begin{equation}
	\hat a_I(z)=\hat U_I^\dagger(z)\hat a\,\hat U_I(z),\qquad
	\hat b_I^\dagger(z)=\hat U_I^\dagger(z)\hat b^\dagger\,\hat U_I(z).
\end{equation}
Using $[\hat a,\hat K_0]=\tfrac12\hat a$, $[\hat a,\hat K_-]=0$, $[\hat a,\hat K_+]=\hat b^\dagger$, one finds
\begin{align}
	i\hbar v\,\partial_z\hat a_I &= \Lambda\bigl(\hat a_I+e^{i\Phi}\hat b_I^\dagger\bigr),\\
	i\hbar v\,\partial_z\hat b_I^\dagger &= -\Lambda\bigl(\hat b_I^\dagger+e^{-i\Phi}\hat a_I\bigr).
\end{align}

\subsection{Rotating frame and conserved norm}

The symmetric rotation $\hat a_R=e^{-i\Phi/2}\hat a_I$, $\hat b_R^\dagger=e^{i\Phi/2}\hat b_I^\dagger$ removes the oscillating phases,
yielding $i\hbar v\,\partial_z(\hat a_R,\hat b_R^\dagger)^T=\mathcal H(z)(\hat a_R,\hat b_R^\dagger)^T$ with
\begin{equation}
	\mathcal H(z)=
	\begin{pmatrix}
		\Delta(z) & \Lambda(z)\\
		-\Lambda(z) & -\Delta(z)
	\end{pmatrix},\quad
	\Delta(z)=\hbar\omega_L(z_0)+\Lambda(z).
\end{equation}
Since $\mathcal H^\dagger\sigma_z=\sigma_z\mathcal H$, the SU(1,1) norm
\begin{equation}
	|\langle\hat a_R\rangle_0|^2-|\langle\hat b_R^\dagger\rangle_0|^2=\text{const},
\end{equation}
where $\langle\cdot\rangle_0\equiv\langle\chi_I(z_0)|\cdot|\chi_I(z_0)\rangle$.
\subsection{Return to the Schr\"odinger picture}

We now trace the sequence of unitary transformations backwards to relate the
conserved SU(1,1) norm to the Schr\"odinger-picture expectation values, which
are directly connected to the extrinsic OAM via
$\mathcal L_{\mathrm{ext}}/\hbar = |\langle\hat a\rangle|^2 - |\langle\hat b^\dagger\rangle|^2$.

Recall the three layers of the construction:
\begin{itemize}
	\item Schr\"odinger picture: state $|\chi(z)\rangle$, fixed operators $\hat a,\hat b^\dagger$.
	\item Interaction picture: $|\chi_I(z)\rangle = \hat U^\dagger(z)|\chi(z)\rangle$,
	fixed operators $\hat a,\hat b^\dagger$.
	\item Heisenberg-in-interaction picture: fixed state $|\chi_I(z_0)\rangle$,
	evolving operators $\hat a_I(z) = \hat U_I^\dagger(z)\,\hat a\,\hat U_I(z)$,
	$\hat b_I^\dagger(z) = \hat U_I^\dagger(z)\,\hat b^\dagger\,\hat U_I(z)$.
\end{itemize}
The rotating-frame operators are $\hat a_R = e^{-i\Phi/2}\hat a_I$,
$\hat b_R^\dagger = e^{i\Phi/2}\hat b_I^\dagger$, and
$\langle\cdot\rangle_0 \equiv \langle\chi_I(z_0)|\cdot|\chi_I(z_0)\rangle$.

Because $[\hat H_0+\Omega\hat{\mathcal L},\hat a] \propto \hat a$ and
$[\hat H_0+\Omega\hat{\mathcal L},\hat b^\dagger] \propto \hat b^\dagger$,
both $\hat a$ and $\hat b^\dagger$ are eigenoperators of the unitary
$\hat U(z)$. Consequently, $\hat U$ and $\hat U^\dagger$ act on them by
simple phase factors:
$\hat U^\dagger\hat a\,\hat U = \hat a\,e^{-i[\Phi+\Phi_1]}$ and
$\hat U\hat a\,\hat U^\dagger = \hat a\,e^{+i[\Phi+\Phi_1]}$,
and similarly for $\hat b^\dagger$ with phase $\Phi_1$.

Unwinding the transformations for $\langle\hat a_R\rangle_0$ step by step:
\begin{align}
	\langle\hat a_R(z)\rangle_0
	&= e^{-i\Phi/2}\,\langle\chi_I(z_0)|\,\hat U_I^\dagger(z)\,\hat a\,\hat U_I(z)\,|\chi_I(z_0)\rangle
	\nonumber\\
	&= e^{-i\Phi/2}\,\langle\chi_I(z)|\,\hat a\,|\chi_I(z)\rangle
	\qquad\bigl(|\chi_I(z)\rangle = \hat U_I(z)|\chi_I(z_0)\rangle\bigr)
	\nonumber\\
	&= e^{-i\Phi/2}\,\langle\chi(z)|\,\hat U(z)\,\hat a\,\hat U^\dagger(z)\,|\chi(z)\rangle
	\qquad\bigl(|\chi_I(z)\rangle = \hat U^\dagger(z)|\chi(z)\rangle\bigr)
	\nonumber\\
	&= e^{-i\Phi/2}\,e^{+i[\Phi+\Phi_1]}\,\langle\chi(z)|\hat a|\chi(z)\rangle
	\nonumber\\
	&= e^{+i[\Phi/2+\Phi_1]}\,\langle\hat a\rangle.
\end{align}

For $\langle\hat b_R^\dagger\rangle_0$ the same steps give
\begin{align}
	\langle\hat b_R^\dagger(z)\rangle_0
	&= e^{+i\Phi/2}\,\langle\chi_I(z_0)|\,\hat U_I^\dagger(z)\,\hat b^\dagger\,\hat U_I(z)\,|\chi_I(z_0)\rangle
	\nonumber\\
	&= e^{+i\Phi/2}\,\langle\chi_I(z)|\,\hat b^\dagger\,|\chi_I(z)\rangle
	\nonumber\\
	&= e^{+i\Phi/2}\,\langle\chi(z)|\,\hat U(z)\,\hat b^\dagger\,\hat U^\dagger(z)\,|\chi(z)\rangle
	\nonumber\\
	&= e^{+i\Phi/2}\,e^{+i\Phi_1}\,\langle\chi(z)|\hat b^\dagger|\chi(z)\rangle
	\nonumber\\
	&= e^{+i[\Phi/2+\Phi_1]}\,\langle\hat b^\dagger\rangle.
\end{align}

Both expectation values carry the \emph{same} overall phase factor
$e^{+i[\Phi/2+\Phi_1]}$.

\clearpage

\section{Time evolution of the extrinsic OAM}

\subsection{Heisenberg equations}

The full Hamiltonian in the symmetric gauge is
\begin{equation}
	\hat H = \frac{\hat{\mathbf p}^2}{2m} + \hat\omega_L\,\hat{\mathcal L}
	+ \frac12 m\,\hat\omega_L^2\,\hat r_\perp^2,
	\qquad \hat\omega_L \equiv \omega_L(\hat z) = \frac{e}{2m}B(\hat z),
\end{equation}
with $\hat{\mathcal L} = \hat x\hat p_y - \hat y\hat p_x$ and
$\hat r_\perp^2 = \hat x^2 + \hat y^2$.

From $\frac{d}{dt}\langle\hat O\rangle = \frac{i}{\hbar}\langle[\hat H,\hat O]\rangle$,
using $[\hat{\mathcal L},\hat x]=i\hbar\hat y$, $[\hat{\mathcal L},\hat y]=-i\hbar\hat x$,
$[\hat{\mathcal L},\hat p_x]=i\hbar\hat p_y$, and $[\hat{\mathcal L},\hat p_y]=-i\hbar\hat p_x$:
\begin{align}
	[\hat H,\hat x] &= \frac{1}{2m}[\hat p_x^2,\hat x] + \hat\omega_L[\hat{\mathcal L},\hat x]
	= -\frac{i\hbar}{m}\hat p_x + i\hbar\hat\omega_L\hat y
	= -\frac{i\hbar}{m}\hat\pi_x
	\quad\Rightarrow\quad
	\frac{d\langle\hat x\rangle}{dt} = \frac{\langle\hat\pi_x\rangle}{m},\\[4pt]
	[\hat H,\hat y] &= -\frac{i\hbar}{m}\hat p_y - i\hbar\hat\omega_L\hat x
	= -\frac{i\hbar}{m}\hat\pi_y
	\quad\Rightarrow\quad
	\frac{d\langle\hat y\rangle}{dt} = \frac{\langle\hat\pi_y\rangle}{m},\\[4pt]
	[\hat H,\hat p_x] &= \hat\omega_L[\hat{\mathcal L},\hat p_x]
	+ \frac12 m\hat\omega_L^2[\hat x^2,\hat p_x]
	= i\hbar\bigl(\hat\omega_L\hat p_y + m\hat\omega_L^2\hat x\bigr)
	\quad\Rightarrow\quad
	\frac{d\langle\hat p_x\rangle}{dt}
	= -\langle\hat\omega_L\hat p_y\rangle - m\langle\hat\omega_L^2\hat x\rangle,\\[4pt]
	[\hat H,\hat p_y] &= -i\hbar\bigl(\hat\omega_L\hat p_x - m\hat\omega_L^2\hat y\bigr)
	\quad\Rightarrow\quad
	\frac{d\langle\hat p_y\rangle}{dt}
	= \langle\hat\omega_L\hat p_x\rangle - m\langle\hat\omega_L^2\hat y\rangle,
\end{align}
with the kinetic (mechanical) momenta
\begin{equation}
	\hat\pi_x = \hat p_x - m\hat\omega_L\hat y,
	\qquad
	\hat\pi_y = \hat p_y + m\hat\omega_L\hat x .
	\label{eq:pidef}
\end{equation}
These equations are exact consequences of $\hat H$; no approximation to the
quantum dynamics has been made.

\subsection{Evolution of the extrinsic OAM}

Differentiating $\mathcal L_{\mathrm{ext}} = \langle\hat x\rangle\langle\hat p_y\rangle
- \langle\hat y\rangle\langle\hat p_x\rangle$ and inserting the Heisenberg equations,
\begin{align}
	\frac{d\mathcal L_{\mathrm{ext}}}{dt}
	&= \frac{\langle\hat\pi_x\rangle}{m}\langle\hat p_y\rangle
	+ \langle\hat x\rangle\bigl(\langle\hat\omega_L\hat p_x\rangle - m\langle\hat\omega_L^2\hat y\rangle\bigr)
	- \frac{\langle\hat\pi_y\rangle}{m}\langle\hat p_x\rangle
	- \langle\hat y\rangle\bigl(-\langle\hat\omega_L\hat p_y\rangle - m\langle\hat\omega_L^2\hat x\rangle\bigr) \nonumber\\[4pt]
	&= \frac1m\bigl[\langle\hat\pi_x\rangle\langle\hat p_y\rangle
	- \langle\hat\pi_y\rangle\langle\hat p_x\rangle\bigr]
	+ \bigl[\langle\hat x\rangle\langle\hat\omega_L\hat p_x\rangle
	+ \langle\hat y\rangle\langle\hat\omega_L\hat p_y\rangle\bigr] \nonumber\\
	&\quad + m\bigl[\langle\hat y\rangle\langle\hat\omega_L^2\hat x\rangle
	- \langle\hat x\rangle\langle\hat\omega_L^2\hat y\rangle\bigr].
\end{align}
The kinematic bracket is then expanded via \eqref{eq:pidef}.  The purely
canonical parts cancel trivially, while the gauge-potential contributions give
\begin{equation}
	\frac1m\bigl[\langle\hat\pi_x\rangle\langle\hat p_y\rangle
	- \langle\hat\pi_y\rangle\langle\hat p_x\rangle\bigr]
	= -\bigl[\langle\hat\omega_L\hat x\rangle\langle\hat p_x\rangle
	+ \langle\hat\omega_L\hat y\rangle\langle\hat p_y\rangle\bigr].
\end{equation}
Inserting this result back and grouping terms with the same operator content
yields the exact intermediate form
\begin{equation}\label{eq:Lext_deriv}
	\begin{aligned}
		\frac{d\mathcal L_{\mathrm{ext}}}{dt}
		&= m\bigl[\langle\hat y\rangle\langle\hat\omega_L^2\hat x\rangle
		- \langle\hat x\rangle\langle\hat\omega_L^2\hat y\rangle\bigr] \\
		&\quad + \bigl[\langle\hat x\rangle\langle\hat\omega_L\hat p_x\rangle
		- \langle\hat\omega_L\hat x\rangle\langle\hat p_x\rangle\bigr] \\
		&\quad + \bigl[\langle\hat y\rangle\langle\hat\omega_L\hat p_y\rangle
		- \langle\hat\omega_L\hat y\rangle\langle\hat p_y\rangle\bigr].
	\end{aligned}
\end{equation}

To expose the role of correlations we introduce the covariance
$\mathrm{Cov}(\hat o_1,\hat o_2)=\langle\hat o_1\hat o_2\rangle
-\langle\hat o_1\rangle\langle\hat o_2\rangle$.
Writing each average as a covariance plus a factorized contribution,
\begin{align}
	\langle\hat x\rangle\langle\hat\omega_L\hat p_x\rangle
	- \langle\hat\omega_L\hat x\rangle\langle\hat p_x\rangle
	&= \langle\hat x\rangle\bigl[\mathrm{Cov}(\hat\omega_L,\hat p_x)
	+\langle\hat\omega_L\rangle\langle\hat p_x\rangle\bigr]
	-\bigl[\mathrm{Cov}(\hat\omega_L,\hat x)
	+\langle\hat\omega_L\rangle\langle\hat x\rangle\bigr]\langle\hat p_x\rangle
	\nonumber\\
	&= \langle\hat x\rangle\,\mathrm{Cov}(\hat\omega_L,\hat p_x)
	- \langle\hat p_x\rangle\,\mathrm{Cov}(\hat\omega_L,\hat x),
\end{align}
the pieces $\langle\hat x\rangle\langle\hat\omega_L\rangle\langle\hat p_x\rangle$
cancelling pairwise.  The remaining two brackets are transformed in the same
way.  For the $m[\cdots]$ part one expands
\begin{align}
	m\langle\hat y\rangle\langle\hat\omega_L^2\hat x\rangle
	&= m\langle\hat y\rangle\bigl[\mathrm{Cov}(\hat\omega_L^2,\hat x)
	+\langle\hat\omega_L^2\rangle\langle\hat x\rangle\bigr],\\
	m\langle\hat x\rangle\langle\hat\omega_L^2\hat y\rangle
	&= m\langle\hat x\rangle\bigl[\mathrm{Cov}(\hat\omega_L^2,\hat y)
	+\langle\hat\omega_L^2\rangle\langle\hat y\rangle\bigr],
\end{align}
and the terms containing $\langle\hat\omega_L^2\rangle$ cancel
identically because $\langle\hat y\rangle\langle\hat x\rangle
-\langle\hat x\rangle\langle\hat y\rangle=0$.  Collecting everything,
\begin{equation}\label{eq:Lext_cov}
	\begin{aligned}
		\frac{d\mathcal L_{\mathrm{ext}}}{dt}
		&= m\bigl[\langle\hat y\rangle\,\mathrm{Cov}(\hat\omega_L^2,\hat x)
		- \langle\hat x\rangle\,\mathrm{Cov}(\hat\omega_L^2,\hat y)\bigr] \\
		&\quad + \bigl[\langle\hat x\rangle\,\mathrm{Cov}(\hat\omega_L,\hat p_x)
		- \langle\hat p_x\rangle\,\mathrm{Cov}(\hat\omega_L,\hat x)\bigr] \\
		&\quad + \bigl[\langle\hat y\rangle\,\mathrm{Cov}(\hat\omega_L,\hat p_y)
		- \langle\hat p_y\rangle\,\mathrm{Cov}(\hat\omega_L,\hat y)\bigr].
	\end{aligned}
\end{equation}
\clearpage

\section{First-principles numerical simulations}

\subsection{Propagation algorithm and initial-state wavefunction}

We develop a numerical solver for the three-dimensional time-dependent magnetic Schr\"odinger equation for a scalar electron
\begin{equation}
i\hbar\partial_t\psi=\hat{H}\psi,
\quad
\hat{H}=\frac{\hat{\bm\pi}^2}{2m}=-\frac{\hbar^2}{2m}\laplacian-i\hbar\frac{e}{m}({\bf A}{\bm\cdot}\grad)+\frac{e^2}{2m}{\bf A}^2,
\end{equation}
provided the Coulomb gauge condition $\nabla\!\bm{\cdot}\!\mathbf{A}=0$ is satisfied. Then we perform first-principles simulations of single-electron wavefunctions in coherent vortex states.

In our code, we employ the Strang-splitting scheme temporal discretization, which is third-order accurate and symplectic, and the Fourier pseudo-spectral method for spatial discretization, which is spectrally accurate, based upon the FFT (Fast Fourier Transform) techniques. For correctly treating the paramagnetic operator $(\mathbf{A}\!\bm{\cdot}\!\nabla)$, the so-called semi-Lagrangian method is adopted in combination with 5th-order spline interpolation, which is an unconditionally stable technique commonly used in dealing with advection problems. This design strikes an optimal balance between computational efficiency and solution accuracy, making it a robust tool for solving various Schrödinger-type wave equations. Additionally, all the expectation values are computed via trapezoidal quadrature that gives second-order accurate results.

The numerical scheme for each time-step that we constructed, in theory, is formulated as
\begin{subequations}
	\begin{align}
		&\psi_1(\mathbf{r})=\mathopen{\mathcal{F}^{-1}}\left[\exp\left[\frac{\hbar\Delta{t}}{4im}\mathbf{k}^2\right]\mathopen{\mathcal{F}}\left[\psi(t,\mathbf{r})\right]\right]\\
		&\psi_2(\mathbf{r})=\exp\left[\frac{e^2\Delta{t}}{4im\hbar}\mathbf{A}^2(\mathbf{r})\right]\psi_1(\mathbf{r})\\
		&\psi_3(\mathbf{r})=\psi_2(\mathbf{s}(t))~~\text{for}~
		\left\{\,\begin{aligned}
			&\frac{\mathrm{d}\mathbf{s}(t)}{\mathrm{d}t}
			=\frac{-e}{m}\mathbf{A}(t,\mathbf{s}(t)),\\
			&\mathbf{s}(t\!+\!\Delta{t})=\mathbf{r}
		\end{aligned}\,\right\}\\
		&\psi_4(\mathbf{r})=\exp\left[\frac{e^2\Delta{t}}{4im\hbar}\mathbf{A}^2(\mathbf{r})\right]\psi_3(\mathbf{r})\\
		&\psi({t\!+\!\Delta{t}},\mathbf{r})
		\approx\mathopen{\mathcal{F}^{-1}}\left[\exp\left[\frac{\hbar\Delta{t}}{4im}\mathbf{k}^2\right]\mathopen{\mathcal{F}}\left[\psi_4(\mathbf{r})\right]\right].
	\end{align}
\end{subequations}
Here $\mathbf{r}\!=\!(x,y,z)^T$ and $\mathbf{k}\!=\!(k_x,k_y,k_z)^T$ mean rectangular grid-points in discrete 3-dimensional real space and the related wavenumber space in the code. The operators ${\mathcal{F}}$ and ${\mathcal{F}^{-1}}$ denote the Fourier transformation from $\mathbf{r}$-space to $\mathbf{k}$-space and its inversion---FFT algorithms in practice. The third line represents a solution to the advection sub-problem $\partial_t\psi+\tfrac{e}{m}(\mathbf{A}\!\bm{\cdot}\!\nabla)\psi=0$ by semi-Lagrangian method. Essentially, this scheme primarily involves a traceback procedure (solving the set of ODEs) followed by a coordinate mapping (to update the wavefunction) for each time-step. In our code, the traceback procedure computes the corresponding interpolation coordinate $\mathbf{s}$ for each spatial grid-point $\mathbf{r}$. Then, the coordinate mapping $\psi(t+\Delta{t},\mathbf{r})=\psi(t,\mathbf{s})$ implies that value of $\psi(t+\Delta{t})$ at grid-point $\mathbf{r}$ equals the estimated value of $\psi(t)$ at the associated interpolation point $\mathbf{s}$.

At $t=0$, the electron wavefunction is initiated with a Laguerre-Gaussian wavefront and a helical phase plus a Gaussian longitudinal envelope, centered at $\mathbf{r}_0$, with mean velocity $v_0$. The initial wavefunction is formulated as
\begin{equation}\label{eq:wavefunc}
	\begin{aligned}
		&\vb{r}'=\vb{r}-\vb{r}_0,~r'_\perp=\sqrt{x'^2+y'^2},~\theta'=\arctan[y'/x'],\\
		&\psi_0=\mathcal{N}_\perp\frac{ (r'_\perp)^{\abs{\ell}}}{w_\perp^{\abs{\ell}+1}}\mathopen{L_n^{\abs{\ell}}}\left[\frac{(r'_\perp)^2}{w_\perp^2}\right]\exp[-\frac{(r'_\perp)^2}{2w_\perp^2}+{\rm{i}}\ell\theta']\\
		&\qquad\times\frac{\mathcal{N}_z}{\sqrt{w_z}}\exp[-\frac{(z')^2}{2w_z^2}-\frac{m}{{\rm{i}}\hbar}v_0z']
		\exp[\frac{e}{{\rm{i}}\hbar}\Lambda(\vb{r})],\\
		&\Lambda(\vb{r})
		=\int_{\vb{r}_0}^{\vb{r}}{\rm d}\vb{s}\cdot\vb{A}(\vb{s})
		=\tfrac{1}{2}B(z)[x_0y-xy_0].
	\end{aligned}
\end{equation}
Here $e\Lambda(\vb{r})/\hbar$ is a local gauge phase. $w_\perp$ and $w_z$ are respectively the radial and axial characteristic widths of the wavepacket. $L_n^{\abs{\ell}}[X]$ means the generalized Laguerre-Gaussian polynomial associated to radial index $n$ and topological charge $\ell$. $\mathcal{N}_{\perp}$ and $\mathcal{N}_{z}$ represent the transverse and longitudinal normalization factors, respectively.

\subsection{Supplemental simulation results}

\begin{figure}[h]
	\centering
	\includegraphics[width=0.6\linewidth]{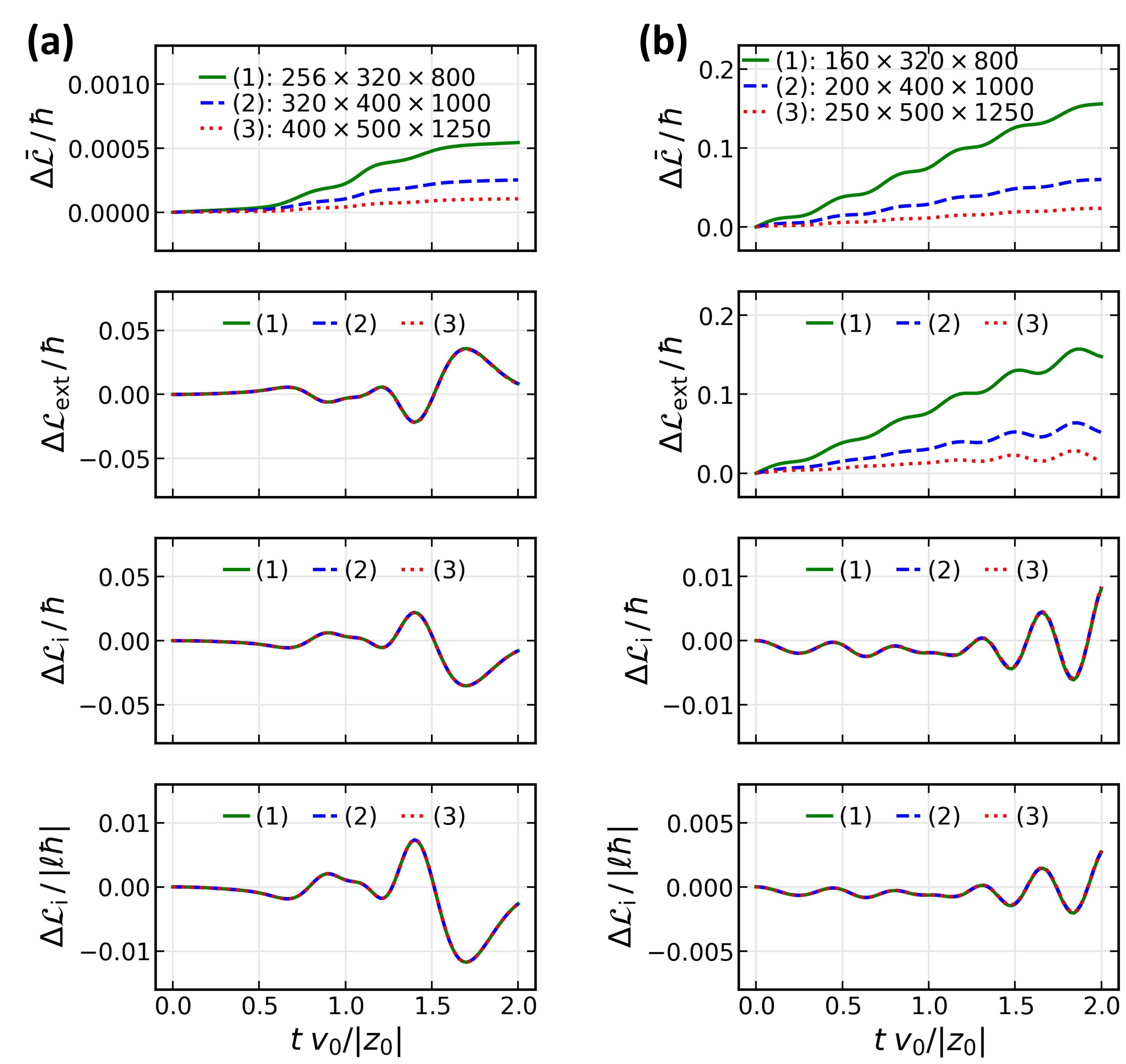}
	\caption{Mesh-refinement tests for settings (a) associated to the case $a=40\,w_m$ in Fig.~2 and (b) the case $y_0=+6\,w_m$ in Fig.~3. For each column, the figure legend marks the spatial resolution $N_x\times N_y\times N_z$ of different cases with identical simulation domain. Note that we adopt a $320\times400\times1000$-resolution mesh for the simulations depicted in Fig.~2 and $200\times400\times1000$-resolution for those associated to Fig.~3 of the main context. For cases in the left column (a), a vortex wavepacket with topological charge $\ell=+3$, longitudinal speed $v_0=160w_m/(6T_c)$, characteristic width $w_\perp=2\,w_m$ and length $w_z=4\,w_m$, initially centering at $(0,+2w_m,-80w_m)$, is released in a Glaser field characterized by $a=40\,w_m$. For cases in the right column (b), a vortex wavepacket with topological charge $\ell=+3$, characteristic width $w_\perp=1\,w_m$ and length $w_z=4\,w_m$, initially centering at $(0,+6w_m,-80w_m)$, is released in a Glaser field characterized by $a=160\,w_m$ with longitudinal speed $v_0=160w_m/(6T_c)$ the topological charge is $\ell=+3$ and the centroid speed is $v_0=160w_m/(6T_c)$. All these cases apply the peak field strength $B_0=\SI{0.02}{T}$ which corresponds to characteristic beam-width $w_m=\sqrt{2\hbar/(eB_0)}=\SI{0.256}{\micro\meter}$ and cyclotron period $T_c=2\pi\,m/(eB_0)=\SI{1.786}{ns}$. Each case experiences 100 time-steps.}
	\label{fig:scan-res}
\end{figure}

Figure~\ref{fig:scan-res} shows the results of grid-refinement tests for the settings corresponding to the case $a=40\,w_m$ in Fig.~2(b) and  the case $y_0=+6\,w_m$ in Fig.~3(b) of the main context, which exhibit the largest deviations in $\mathcal{L}_\text{i}$ respectively. In our simulations, the rectangular simulation domain is discretized on a regular $N_x\times N_y\times N_z$-resolution grid mesh. In the main context, we adopt a $320\times400\times1000$-resolution mesh for the simulations shown in Fig.~2 [and also Movies S1--S3] and $200\times400\times1000$-resolution for those presented to Fig.~3  [and also Movies S4--S6]. Taking these meshes as the baseline resolutions, denser ($\times1.25$) and coarser ($\times0.8$) meshes lead to no appreciable change in $\Delta\mathcal{L}_\text{i}$, thereby confirming numerical convergence of the computed intrinsic canonical OAM $\mathcal{L}_\text{i}$ shown in Fig.~2(b) and Fig.~3(b) of the main context. These results indicate that the observed percent-level fluctuations $\Delta\mathcal{L}_{\text{i}}$ are predominantly physical rather than numerical in origin.

Furthermore, since the axial component of canonical OAM is strictly conserved, the variation of its expectation value relative to the initial value i.e. $\Delta\bar{\mathcal{L}}=\ev{\hat{\mathcal{L}}}{\psi(t)}-\ev{\hat{\mathcal{L}}}{\psi(0)}$ is purely numerical error. In Fig.~\ref{fig:scan-res}, the first row shows that $|\Delta\bar{\mathcal{L}}|$ systematically decreases with mesh refinement, demonstrating convergence of the numerical scheme.

\begin{figure}[h]
	\centering
	\includegraphics[width=1\linewidth]{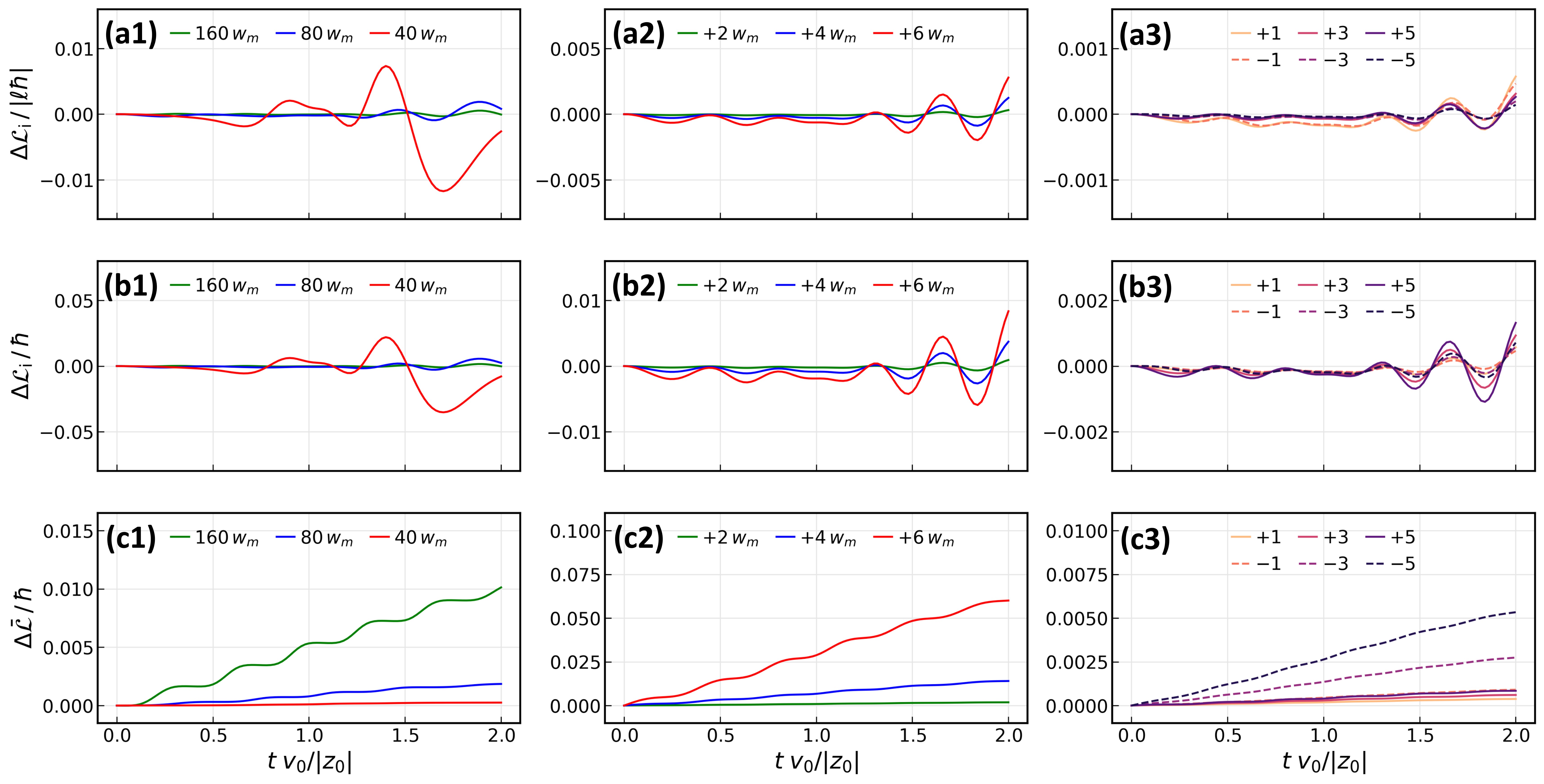}
	\caption{(a*) Relative variation of intrinsic canonical OAM,  (b*) absolute variation of intrinsic canonical OAM, and (c*) absolute variation of canonical OAM for distinct (*1) $a$, (*2) $y_0$, and (*3) $\ell$. For the cases in column (*1), the initial centroid is fixed at $(0,+2,-80)\,w_m$ within Glaser fields of distinct magnetic variation distance $a\in[40,160]\,w_m$, transverse and longitudinal spreads of the initial wavepacket are set to be $w_\perp=2\,w_m$ and $w_z=4\,w_m$ respectively, the topological charge is $\ell=+3$ and the centroid speed is $v_0=160w_m/(6T_c)$. In column (*2), in the same Glaser field characterized by $a=160\,w_m$, the centroid is displaced at different transverse offset $y_0\in[+2,+6]\,w_m$ while the longitudinal coordinate of the initial centroid is $z_0=-80\,w_m$, transverse and longitudinal spreads are $w_\perp=1\,w_m$ and $w_z=4\,w_m$ respectively, the topological charge is $\ell=+3$ and the centroid speed is $v_0=160w_m/(6T_c)$. In column (*3), the settings, except $\ell\in[-5,+5]$, are the same as those of case $y_0=+2\,w_m$ in (*2). In these cases, we use $B_0=\SI{0.02}{T}$ which corresponds to characteristic beam-width $w_m=\sqrt{2\hbar/(eB_0)}=\SI{0.256}{\micro\meter}$ and cyclotron period $T_c=2\pi\,m/(eB_0)=\SI{1.786}{ns}$. The simulations shown in column (*1) are conducted within $200\times 400\times 1000$-resolution mesh on $20\times 40\times 200\,w_m^3$ rectangular space domain, and in columns (*2--3) $200\times 200\times 1000$-resolved $20\times 20\times 200\,w_m^3$ space is used. Each case experiences 100 time-steps.}
	\label{fig:oam}
\end{figure}

Figure~\ref{fig:oam} presents numerical results of $\Delta\bar{\mathcal{L}}$ and $\Delta{\mathcal{L}}_{\text{i}}$ against different characteristic magnetic variation length $a$ of Glaser fields, initial transverse offset $y_0$ of the electron wavepacket centroid, and initial topological charge $\ell$ of the vortex state, separately. In rows (b*--c*) of Fig.~\ref{fig:oam}, maximum values of $\Delta{\mathcal{L}}_{\text{e}}$ and $\Delta\bar{\mathcal{L}}$ are roughly on the same order of magnitude. However, in these simulations, the change of $\mathcal{L}_{\text{i}}$ relative to $\ell\hbar$, generally with relative variation $\lesssim1\%$ [Figs.~\ref{fig:oam}(a)], can be smaller than the instantaneous magnitude of $\Delta\bar{\mathcal{L}}$, manifesting excellent robustness of intrinsic canonical OAM.

The main observations may be summarized as follows:

(i)
Reducing the characteristic variation length \(a\) increases the amplitude of the intrinsic-OAM oscillations.
Nevertheless, the maximum relative deviation remains at the percent level for all parameters considered.

(ii)
The variation amplitude grows with the transverse offset of the wavepacket centroid, yet remains below \(0.3\%\) for displacements of several characteristic beam widths.

(iii)
For \(\ell\in[-5,+5]\), the maximum relative deviation stays below \(0.1\%\), and no systematic scaling with the topological charge is observed.

\clearpage

\section{Description of Supplemental Videos}

The movies below show the time evolution of the real-space wavefunction of a vortex electron. The orthogonal planes intersect at the instantaneous wavepacket centroid, $\bar{\bf r}=\ev{\hat{\bf r}}{\psi}$. The hue represents the phase of the complex wavefunction, $\theta=\arg[\psi]$, while the lightness is proportional to the normalized probability density, $\rho=|\psi|^2$.

\begin{description}
\item[Movie S1]  Wavefunction of vortex electron depicted in Fig.~2(d).
\item[Movie S2]  Wavefunction of vortex electron depicted in Fig.~2(e).
\item[Movie S3]  Wavefunction of vortex electron depicted in Fig.~2(f).
\item[Movie S4]  Wavefunction of vortex electron depicted in Fig.~3(d).
\item[Movie S5]  Wavefunction of vortex electron depicted in Fig.~3(e).
\item[Movie S6]  Wavefunction of vortex electron depicted in Fig.~3(f).
\end{description}